\documentclass[twocolumn,prl,aps,showpacs]{revtex4}
\usepackage{amssymb}

\usepackage{amsfonts}
\usepackage{amsmath}
\usepackage{epsfig}
\usepackage{array}
\usepackage{rotating,booktabs}

\begin{document}

\title{{Quantum dynamics of atomic Rydberg excitation in strong laser fields }

\footnotetext{$^{*}$ These authors contribute equally to this work.}
\footnotetext{$^{\S}$xuhf@jlu.edu.cn}
\footnotetext{$^{\dag}$wbecker@mbi-berlin.de}
\footnotetext{$^{\ddag}$chen$\_$jing@iapcm.ac.cn}
}

\author{ Shilin Hu$^{1*}$, Xiaolei Hao$^{2,*}$, Hang Lv$^{3,4}$, Mingqing
Liu$^{5}$, Tianxiang Yang$^{3,4}$, Haifeng Xu$^{3,4,\S}$, Mingxing
Jin$^{3,4}$, Dajun Ding$^{3,4}$, Qianguang Li$^{6}$, Weidong
Li$^{2}$, Wilhelm Becker$^{7,8,\dag}$, and Jing Chen$^{5,9,\ddag}$
}

\affiliation{$^{1}$ Laboratory of Quantum Engineering and Quantum
Metrology, School of Physics and Astronomy, Sun Yat-Sen University
(Zhuhai Campus), Zhuhai 519082, China.\\
$^{2}$ Institute of Theoretical Physics and Department of Physics,
Shanxi University, 030006 Taiyuan, China. \\
 $^{3}$ Institute of
atomic and molecular physics, Jinlin University, Changchun 130012,
China.\\
 $^{4}$ Jilin Provincial Key Laboratory of Applied Atomic
and Molecular Spectroscopy, Jilin University, Changchun 130012,
China.\\
$^{5}$ Institute of Applied Physics and Computational Mathematics,
P. O. Box 8009, Beijing 100088, China.\\
 $^{6}$
School of Physics and Electronic-information Engineering, Hubei
Engineering University, Xiaogan 432000,
China.\\
$^{7}$ Max-Born-Institut, Max-Born-Strasse 2a, 12489 Berlin,
Germany.\\
 $^8$ National Research Nuclear University MEPhI, 115409
Moscow, Russia.\\
$^{9}$ Center for Advanced Material Diagnostic Technology,
Shenzhen Technology University, Shenzhen 518118, China. \\}

\date{ \today }

\begin{abstract}
Neutral atoms have been observed to survive intense laser pulses
in high Rydberg states with surprisingly large probability. Only
with this Rydberg-state excitation (RSE) included is the picture
of intense-laser-atom interaction complete. Various mechanisms
have been proposed to explain the underlying physics. However,
neither one can explain all the features observed in experiments
and in time-dependent Schr\"{o}dinger equation (TDSE) simulations.
Here we propose a fully quantum-mechanical model based on the
strong-field approximation (SFA). It well reproduces the intensity
dependence of RSE obtained by the TDSE, which  exhibits a series
of modulated peaks. They are due to recapture of the liberated
electron and the fact that the pertinent probability strongly
depends on the position and the parity of the Rydberg state. We
also present measurements of RSE in xenon at 800 nm, which display
the peak structure consistent with the calculations.

\end{abstract}

\maketitle

\section{Introduction}

In strong-field atomic and molecular physics, Rydberg states
attracted much attention in the 1990s
\cite{Freeman1987PRL,Jones1993,Muller1988PRL,Rottke1994PRA} but
thereafter have been ignored for a long time. This is because, in
an intense laser field, the atoms or molecules are so strongly
disturbed that the electrons already in the ground state of the
neutral atom or even the ion can be liberated very efficiently
\cite{ap79,wbecker1,Becker2012RMP}. This appeared to imply that
the weakly bound Rydberg states would lose any significance. Only
recently, however, both theoretical and experimental works
surprisingly found that quite a large portion of neutral atoms in
Rydberg states can survive very strong laser fields
\cite{wang2006CPL,EichmannPRL08}. This has become the object of
elaborate experimental and theoretical studies in the past few
years
\cite{Eichmann2009Nat,NISS2009,voklova2010JETP,sandner2013PRL,ZPIE2017,Eichmann2013PRL,huang2013PRA,Azarm2013JPCS,Eichmann2012PRA,lin2014PRA,Eichmann2015PRL,Piraux2017PRA,Popruzhenko2018}.
Besides atoms, Rydberg state excitation (RSE) has also been
observed in atomic fragments from Coulomb explosion of dimers
\cite{wu2011PRL} and diatomic molecules
\cite{Eichmann2009PRL,Lv2016PRA}.

Theoretically, a semiclassical two-step model has been proposed to
explain the experimental observations
\cite{wang2006CPL,EichmannPRL08,Eichmann2009Nat,NISS2009}. At
first, the electron is pumped into a continuum state from its
initial bound state by the external field via tunneling
\cite{keldysh,ADK}. The subsequent description of the
electron in the continuum follows completely classical lines
\cite{Schafer1993,Corkum1993PRL}. Owing to the presence of the
attractive Coulomb field of the ion, the electron may be left with
negative energy at the end of the laser pulse, which corresponds
to capture into a Rydberg state \cite{wang2006CPL}, also known as
frustrated tunneling ionization (FTI) \cite{EichmannPRL08}. An
alternative theoretical approach proceeds via the solution of the
time-dependent one-electron Schr\"odinger equation (TDSE)
\cite{voklova2010JETP,lin2014PRA,ZPIE2017}. This includes, of
course, all quantum effects, but it is difficult to extract from
it much physical insight into the details of the dynamics.
In fact, two mechanisms have been proposed to explain the peak
structure in the intensity dependence of the RSE population, which is
clearly visible in the TDSE calculation. One is the
Freeman-resonance perspective in which the Rydberg states are
populated via a resonant multiphoton transition
\cite{voklova2010JETP,Freeman1987PRL}. The other one suggests that
RSE is just the continuation of above-threshold ionization (ATI)
to negative energies in the Rydberg quasi-continuum
\cite{lin2014PRA}. On the other hand, closer inspection shows that
the peaks in the RSE population as a function of intensity
alternate between comparatively low and high yield, which was not
addressed in these papers \cite{voklova2010JETP,lin2014PRA}.
Apparently, the mechanism via Freeman resonances is unable to
explain this effect, and the existence of peak structures is
clearly beyond the scope of the afore-mentioned semiclassical
model.

\begin{figure}[htb]
\centering\vspace{-0in}\includegraphics[width=3.6 in]{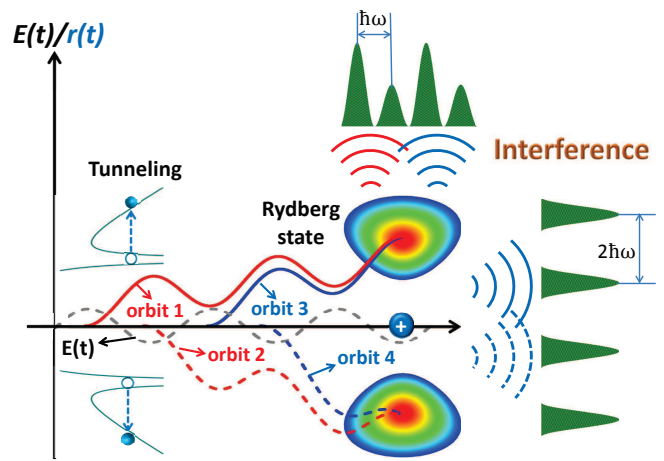}
\caption{(color online). The Rydberg-state excitation process: The
dashed gray curve denotes the laser electric field. The valence
electron is liberated through tunneling in one of the various
optical cycles of the pulse, whereupon it  evolves in the laser
field. The electrons captured into a certain Rydberg state in one
direction (orbit 1 \& orbit 3) or the other (orbit 2 \& orbit 4)
have different phases and will interfere, which leads to a peak
structure in the intensity dependence with interval of $\Delta
U_p=\hbar\omega$. Interference between the two directions leads to
a peak structure with interval of $\Delta
U_p=2\hbar\omega$ (see text for details).}\label{fig1}%
\end{figure}

In this paper, we formulate a quantum-mechanical model of the RSE
process and compare its results with experiments and TDSE
calculations (which are in a sense almost equivalent to real
experimental data). Our quantum model well reproduces all the
features observed in a TDSE simulation, including the dependence
on the parities of the initial and final states and the
just-mentioned fact that the peaks in the RSE intensity dependence
alternate in height. The model is also applied to the
investigation of RSE of He and Xe atoms. For He, we reproduce the
experimentally observed distributions of the principal quantum
number of the Rydberg states \cite{EichmannPRL08,Eichmann2013PRL}.
For Xe, we present to our knowledge the first measurements of RSE
for 800-nm laser wavelength that display the peak structure. Our
work provides a quantum picture of RSE in intense laser fields:
first, the electron is pumped by the laser field into a continuum
state. Subsequently, it evolves in the external field. Some of the
electrons may be coherently captured into Rydberg states. The
probability of the capture process strongly depends on the
location and the parity of the final Rydberg state, resulting in
all the quantum features observed. A schematic picture of the RSE
process is
presented in Fig. \ref{fig1}.\\

\section{Quantum model}

The transition amplitude of the capture process is given by (atomic units
$m=\hbar=e=1$ are used)
\begin{equation}
M_{nlm}=\lim_{t\rightarrow\infty, t'\rightarrow -\infty}\big<\Psi_{nlm}(t)|U(t,t')|
\Psi_g(t')\big>, \label{eq1}
\end{equation}
where $|\Psi_{nlm}\left(  t\right)  \rangle$ denotes a field-free Rydberg
state with principal quantum number $n$, angular-momentum
quantum number $l$, and magnetic quantum number $m$. $U(t,t')$ denotes the time-evolution
operator with the Coulomb and the laser fields \cite{wbecker1},
and $\Psi_{g}(\mathbf{r},t)=e^{iI_{p}t}\phi_{g}(\mathbf{r})$ indicates the wave
function of the field-free ground state with the ionization potential
$I_p$. It is assumed that the laser field is turned off in the distant
past and future. The time-evolution operator $U(t,t')$ satisfies the
Dyson equation
\begin{equation}
U(t,t')=U_0(t,t')-i\int_{t'}^{t}d\tau U(t,\tau)H_{I}(\tau)U_{0}(\tau,t'),
\label{eq2}
\end{equation}
where $U_{0}(t,t')$ is the time-evolution operator for only the Coulomb
field \cite{wbecker1} and (in length gauge) $H_{I}(t)=\textbf{r}\cdot \textbf{E}(t)$.
With the help of Eq. (\ref{eq2}), the transition amplitude  (\ref{eq1})
can be rewritten as
\begin{align}\label{eq3}
M_{nlm}=&\lim_{t\rightarrow\infty}\Big\{\big<\Psi_{nlm}(t)|\Psi_g(t)\big> \nonumber\\
&-i\int_{-\infty}^{t}
d\tau\big<\Psi_{nlm}(t)|U(t,\tau)H_{I}(\tau)|\Psi_g(\tau)\big>\Big\}.
\end{align}
Here, the first term on the right-hand side is zero due to the orthogonality
of the Coulomb eigenstates. For the second term, we use a different form of
the Dyson equation,
\begin{equation}
U(t,t')=U_V(t,t')-i\int_{t'}^{t}d\tau U(t,\tau)VU_V(\tau,t'),
\label{eq4}
\end{equation}
where $V$ is the binding (Coulomb) potential and $U_V(t,t')$ indicates the Volkov time-evolution operator
\begin{equation}
U_{V}(t,t')=\int d^3\mathbf{k}|\Psi_\mathbf{k}^{(V)}(t)\rangle\langle\Psi_\mathbf{k}^{(V)}(t')|,
\label{eq5}
\end{equation}
which we have expanded in terms of the Volkov states with the wave functions
\begin{align}
\Psi_{\mathbf{k}}^{(V)}(\mathbf{r}, t)=&\frac{1}{ (2\pi)^{3/2}}\exp\Big\{i(\mathbf{k}+\mathbf{A}(t))
\cdot\mathbf{r}\nonumber\\
&-\frac{i}{2}\int^{t}dt'(\mathbf{k}+\mathbf{A}(t^{\prime}))^{2}\Big\},
\label{eq6}
\end{align}
where $\mathbf{k}$ denotes the asymptotic (drift) momentum. The coordinate-space
representation of the Volkov time-evolution operator is
\begin{equation}\label{eq7}
U_V(\mathbf{r}t,\mathbf{r}'t')=i\left(\frac{i}{2\pi(t-t')}\right)^\frac32 e^{-iS(\mathbf{r}t,\mathbf{r}'t')}
\end{equation}
with
\begin{align}\label{volkovaction}
S(\mathbf{r}t,\mathbf{r}'t')=&\mathbf{A}(t)\cdot\mathbf{r}-\mathbf{A}(t')\cdot\mathbf{r}'+\frac12\int^t_{t'}d\tau\mathbf{A}^2(\tau) \nonumber\\
&-\frac{1}{2(t-t')}\Big[\mathbf{r}-\mathbf{r}'+\int^t_{t'}d\tau\mathbf{A}(\tau)\Big]^2.
\end{align}

Using Eq.~(\ref{eq4}) in Eq.~(\ref{eq3}) we get
\begin{align}\label{eq9}
M_{nlm} =& \lim_{t\rightarrow\infty}(-i)\int_{-\infty}^{t}
d\tau\big<\Psi_{nlm}(t)|U_V(t,\tau)H_{I}(\tau)|\Psi_g(\tau)\big> \nonumber \\
 &+\lim_{t\rightarrow\infty}(-i)^2\int_{-\infty}^{t}d\tau'\int^{\tau'}_{-\infty}d\tau \nonumber \\
& \times\big<\Psi_{nlm}(t)|U(t,\tau')VU_V(\tau',\tau)H_{I}(\tau)|
\Psi_g(\tau)\big>.
\end{align}
This is analogous to the Born expansion of the ionization amplitude for
above-threshold ionization. In that case, the final state is in the continuum
and is approximated by a plane wave so that $\langle\Psi_{nlm}(t)|U_V(t,\tau)$
becomes a Volkov state $\langle\Psi_\mathbf{p}^V(\tau)|$. The first term then
describes the direct electrons and the second term (with the exact propagator
$U(t,t')$ replaced by the Volkov propagator $U_{V}(t,t')$) indicates single rescattering.
For energies below $\approx 2U_p$, the first term is dominant. However, for RSE
the first term in Eq. (\ref{eq5}) goes to zero in the limit of $t\to\infty$. This
is due to the factor of $(t-t')^{-3/2}$ in Eq.~(\ref{eq7}), which accounts
for wave-function spreading, and the fact that the Rydberg state $|\Psi_{nlm}(t)\rangle$
is localized. (Note that the values of $\tau$ in Eq.~(\ref{eq9}) are restricted
to the finite duration of the laser pulse.) Hence, for RSE we only consider
the second term.

In order to make further progress we replace the bra $\langle\Psi_{nlm}(t)|U(t,\tau')$
in the second term  by an approximate field-dressed Rydberg state with the wave function
\begin{equation}
\Psi_{nlm}^d(\mathbf{r},\tau')=\phi_{nlm}(\mathbf{r})e^{-iE_{n}\tau'}e^{i\mathbf{r}%
\cdot\mathbf{A}(\tau')}e^{-i\int_{-\infty}^{\tau'}d\tau
A^{2}(\tau)/2}\label{eq10},
\end{equation}
where $\phi_{nlm}(\mathbf{r})$ is a field-free Rydberg state corresponding to the energy
level $E_{n}=-1/(2n^{2})$.  The principal quantum number, angular quantum number, and
magnetic quantum number are indicated by $n$, $l$, and $m$, respectively.
The field-free Rydberg states are given by
\begin{align}\label{s2}
\nonumber   \phi_{nlm}(\mathbf{r})&= N_{nl}R_{nl}(r)Y_{lm}(\theta,\varphi),\\
\nonumber   N_{nl}&=\frac{(2\kappa_n)^{3/2}}{\Gamma(2l+2)}\sqrt\frac{\Gamma(n+l+1)}{2n\Gamma(n-l)},\\
            R_{nl}(r)&=(2\kappa_nr)^le^{-\kappa_nr}\ _1F_1(-n+l+1,2l+2,2\kappa_nr),
\end{align}
where $\kappa_n=1/n$, $Y_{lm}(\theta,\varphi)$ is a spherical
harmonic function, and $_1F_1(x,y,z)$ is the confluent
hypergeometric function. The approximation (\ref{eq10}), called
the Coulomb-Volkov state, has been frequently used to account for
the Coulomb-field in noncontinuum states; see
\cite{Reiss1977,Jain1978} and many later references.
The field-dressed state (\ref{eq10}) does not exactly satisfy the Schr\"odinger equation. Namely, we have
\begin{align}%
i\frac{\partial}{\partial t}\Psi_{nlm}^d(\mathbf{r},t)=&[E_{n}\phi_{nlm}(\textbf{r})+\frac{A^{2}(t)}{2}\phi_{nlm}(\textbf{r})+\mathbf{r}%
\cdot\mathbf{E}(t)\phi_{nlm}(\textbf{r})]  \nonumber \\
&\times e^{-iE_{n}t}e^{i\mathbf{r}\cdot\mathbf{A}(t)}e^{-i\int_{0}^{t}\frac{A^{2}(\tau)}{2}d\tau},
\label{eq12}%
\end{align}
and
\begin{align}%
& [-\frac{1}{2}\nabla^{2}\!-\!\frac{1}{r}\!+\!\mathbf{r}\cdot\mathbf{E}(t)]\Psi
_{nlm}^d(\mathbf{r},t)\!=\![E_{n}\phi_{nlm}(\textbf{r})\!-\!i\nabla\phi_{nlm}(\mathbf{r})\cdot\mathbf{A}%
(t) \nonumber \\
&+\!\frac{A^{2}(t)}{2}\phi_{nlm}(\textbf{r})\!+\!\mathbf{r}\cdot\mathbf{E}
(t)\phi
_{nlm}(\textbf{r})]e^{-iE_{n}t}e^{i\mathbf{r}\cdot\mathbf{A}(t)}e^{-i\int_{0}^{t}%
\frac{A^{2}(\tau)}{2}d\tau},
\label{eq13}%
\end{align}
so that
\begin{align}
&\left[i\frac{\partial}{\partial t}+\frac12\mbox{\boldmath$\nabla$}^2 + \frac1r - \mathbf{r}\cdot\mathbf{E}(t)\right]\Psi_{nlm}^d(\mathbf{r},t) \nonumber \\
&=ie^{-iE_{n}t}e^{i\mathbf{r}\cdot\mathbf{A}(t)}e^{-i\int_{0}^{t}
\frac{A^{2}(\tau)}{2}d\tau}\mathbf{A}(t)\cdot\mbox{\boldmath$\nabla$} \phi_{nlm}(\mathbf{r}).\label{eq14}
\end{align}
Hence, the approximation (10) would be exact if it were not for
the term $-i\nabla\phi_{nlm}(\textbf{r})\cdot A(t)$ on the right-hand side of
Eq.~(\ref{eq14}). A comparison of the four terms of Eq.
(\ref{eq13}) is shown in Table I. It can be seen that the
disturbing term $\nabla\phi_{nlm}(\textbf{r})$ is several orders smaller than
the other three terms in the region where the Rydberg state is
concentrated (see Figs. 3 and 8 in the appendix)
for the times of capture (for details, see the semiclassical analysis
part of the appendix) shown in Fig. \ref{figr1}.
Therefore, Eq. (10) can be considered a good approximation to the
dressed Rydberg state, and our  final approximation to the
Rydberg-capture amplitude is
\begin{align}
M_{nlm}&=\!(-i)^2\!\int^\infty_{-\infty}\!dt\!\int^t_{-\infty}\!dt'\!\langle\Psi^d_{nlm}(t)|VU_V(t,t')H_I(t')|
\Psi_g(t')\rangle \nonumber \\
&=(-i)^2\int_{-\infty}^{\infty}dt\int_{-\infty}^{t}dt^{\prime}\int
d^{3}\mathbf{k}\nonumber\\
&  \times\langle\Psi_{nlm}^d( t) \vert V(\mathbf{r}) |\Psi_{\mathbf{k}
}^{(V)}(  t)  \rangle
\langle\Psi_{\mathbf{k}%
}^{(  V)  }(  t^{\prime})  \vert \mathbf{r'}%
\cdot\mathbf{E}(  t^{\prime})  \vert \Psi_{g}(
t^{\prime})  \rangle.
\label{amplitude}
\end{align}

\begin{figure}[ht]
  \centering
  \includegraphics[scale=0.3]{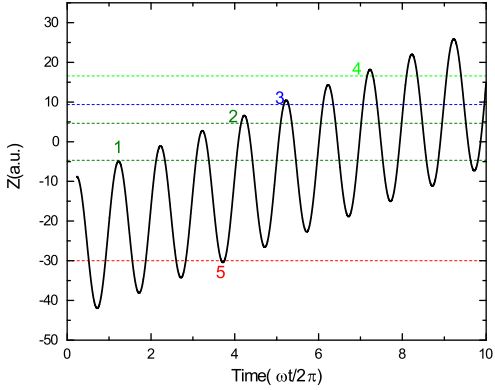}\\
  \caption{A typical trajectory of the electron in the laser pulse with intensity of $1\times 10^{14}$ W/cm$^2$. The numbers indicate
  the instants of time $t$ when the electron is captured by the Rydberg states of $n=6$ of the hydrogen atom with
  different angular momenta. }\label{figr1}
\end{figure}

\begin{table}
\caption{\label{tab:2} Relative amplitudes of the four terms of Eq.
(\ref{eq13}). The quantities $t$ and $l$ indicate the capture moment and the
angular momentum number shown in Fig. \ref{figr1}, respectively. }
\begin{tabular}{cccccccc}\hline\hline
& $t$~~ & $l$~~ & E$_n(n=6)$~~ &
$\frac{\nabla\phi_{nlm}(r)}{\phi_{nlm}(r)}\cdot\mathbf{A}(t)$~~ &
$\frac{A^2(t)}{2}$~~ & $\mathbf{r}\cdot\mathbf{E}(t)$ & \\\hline
&1~~ & 2~~& -0.014~~ & $-2.47\times10^{-7}$~~ & 0.009~~ & 0.20  &
\\\hline &2~~ & 2~~& -0.014~~ & $-9.91\times10^{-7}$~~  & 0.147~~ &
0.35  &\\\hline &3~~ & 3~~& -0.014~~ & $-3.62\times10^{-7}$~~ &
0.102~~ & 0.44  &\\\hline &4~~ & 4~~& -0.014~~ &
$-4.16\times10^{-6}$~~ & 0.129~~ & 0.75  &\\\hline &5~~ & 5~~&
-0.014~~ & $2.21\times10^{-6}$~~  & 0.080~~ & -1.45 &\\\hline\hline
\end{tabular}
\end{table}

With the help of Eqs. (\ref{eq7}), (\ref{eq10}), and the binding
potential $V(r)=-1/r$, the capture amplitude (\ref{amplitude}) has the following form
\begin{equation}\label{Mn}
M_{nlm}\!=\!(-i)^2\!\int_{-\infty}^{\infty}\!dt\!\int_{-\infty}^{t}\!dt^{\prime}\!\int
d^{3}\!\mathbf{k}\!V_{nlm,\mathbf{k}}\!V_{\mathbf{k}g}\!\exp[iS_{n}(t,t^{\prime
},\mathbf{k})],
\end{equation}
with
\begin{align}
V_{\mathbf{k}g}&=\!\frac{1}{(2\pi)^{3/2}}\!\int
d^{3}\!\mathbf{r'}\exp\{-i[  \mathbf{k+A}(  t^{\prime})
] \cdot\mathbf{r'}\} \mathbf{r}'\cdot\mathbf{E}(t') \psi_g(\mathbf{r'})\nonumber\\
&=\!\frac{1}{(2\pi)^{3/2}}\!(-i\frac{\partial}{\partial t'})
\int d ^{3}\mathbf{r'}\exp\{-i[  \mathbf{k+A}(  t^{\prime})
]  \cdot\mathbf{r'}\}  \psi_g(\mathbf{r'}). \label{Vgk}
\end{align}
\begin{equation}\label{Vkn}
V_{nlm,\mathbf{k}}=-\frac{1}{(2\pi)^{3/2}}\int d^3\mathbf{r}\phi^*_{nlm}(\mathbf{r}) \frac{1}{r}\exp(i\mathbf{k}\cdot\mathbf{r}),
\end{equation}
and the action
\begin{equation}\label{sact}
S_{n}\!\left(  t,t^{\prime},\mathbf{k}\right)\!=\!\frac{1}{2}\!\int^{t}_{-\infty}\!d\tau\!\mathbf{A}^{2}\left(\tau\right)\nonumber
  -\frac{1}{2}\!\int_{t^{\prime}}^{t}\!d\tau\!\left[  \mathbf{k+A}\left(
\tau\right)  \right]  ^{2}+E_{n}t+I_{p}t^{\prime}.
\end{equation}
\begin{figure}[htbp]
  \centering
  \includegraphics[scale=0.3]{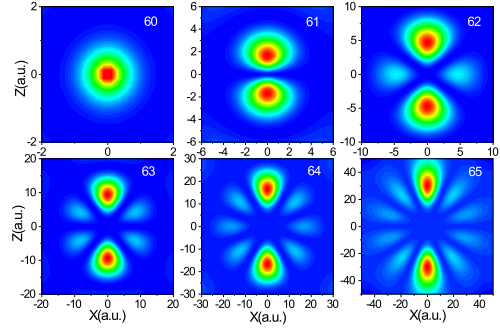}\\
  \caption{Density distribution of the hydrogenic Rydberg state $n = 6$ for different angular quantum numbers $l$.
  The two numbers in each panel correspond to the principal quantum number $n$ and the angular-momentum
  quantum number $l$. }\label{fig33}
\end{figure}

In Fig. \ref{fig33}, the probability densities of the relevant Rydberg states
are plotted  for $n=6$ and  different values of $l$. It can be easily seen that
for $l\neq 0$ the Rydberg state has two centers at $\mathbf{r}_{nlm}^+ \equiv
\mathbf{r}_{nlm}$ and
$\mathbf{r}_{nlm}^-=-\mathbf{r}_{nlm}^+\equiv -\mathbf{r}_{nlm}$.
Accordingly, we decompose the Rydberg-state wave function as
\begin{equation}\label{rr}
  \phi_{nlm}(\mathbf{r})=\phi_{nlm+}(\mathbf{r})+\phi_{nlm-}(\mathbf{r}),
\end{equation}
where the functions $\phi_{nlm\pm}(\mathbf{r})$ are concentrated around $\mathbf{r}=\pm\mathbf{r}_{nlm}$.
Also, $\phi_{nlm-}(-\mathbf{r})=(-1)^l\phi_{nlm+}(\mathbf{r})$, which allows us to write

\begin{align}\label{vnknew}
  V_{nlm,\mathbf{k}} & = -\frac{1}{(2\pi)^{3/2}}\int \frac{d^3\mathbf{r}}{r}\left[\phi^*_{nlm+}(\mathbf{r})e^{i\mathbf{k}\cdot\mathbf{r}}+\phi^*_{nlm-}(\mathbf{r})e^{i\mathbf{k}\cdot\mathbf{r}}\right] \nonumber \\
                  & = -\frac{1}{(2\pi)^{3/2}}\int \frac{d^3\mathbf{r}}{r}\left[\phi^*_{nlm+}(\mathbf{r})e^{i\mathbf{k}\cdot\mathbf{r}}+\phi^*_{nlm-}(-\mathbf{r})e^{-i\mathbf{k}\cdot\mathbf{r}}\right] \nonumber \\
                  &=-\frac{1}{(2\pi)^{3/2}}\int \frac{d^3\mathbf{r}}{r}\phi^*_{nlm+}(\mathbf{r})\left[e^{i\mathbf{k}\cdot\mathbf{r}}+(-1)^l e^{-i\mathbf{k}\cdot\mathbf{r}}\right].
\end{align}
With the abbreviation
$\phi_{nlm+}(\mathbf{r})=\phi_{nlm+}(\mathbf{r}-\mathbf{r}_{nlm}+\mathbf{r}_{nlm})\equiv\tilde{\phi}_{nlm}(\rho)$, where $\rho=\mathbf{r}-\mathbf{r}_{nlm}$,
we end up with
\begin{align}
V_{nlm,\mathbf{k}}&=-\frac{1}{(2\pi)^{3/2}}\int d^3\bm{\rho}
\frac{1}{|\bm{\rho}+\mathbf{r}_{nlm}|}\tilde{\phi}_{nlm}^*(\bm{\rho})\Big[e^{i\mathbf{k}\cdot\bm{\rho}}e^{i\mathbf{k}\cdot\mathbf{r}_{nlm}}\nonumber \\
&+(-1)^le^{-i\mathbf{k}\cdot\bm{\rho}}e^{-i\mathbf{k}\cdot\mathbf{r}_{nlm}}\Big].\label{vnknew'}
\end{align}
We can now proceed with the saddle-point evaluation as it is usually done by
including the fast exponential dependence generated by the factors $\exp(\pm i\mathbf{k}\cdot\mathbf{r}_{nlm})$
into the action while treating the remaining $\mathbf{k}$ dependence as slow.
This means that we replace
\begin{align}
V_{nlm,\mathbf{k}}e^{iS_n(t,t',\mathbf{k})}&=V_{nlm,\mathbf{k}+}e^{i\big[S_n(t,t',\mathbf{k})+\mathbf{k}\cdot\mathbf{r}_{nlm}\big]} \nonumber \\
&+(-1)^lV_{nlm,\mathbf{k}-}e^{i\big[S_n(t,t',\mathbf{k})-\mathbf{k}\cdot\mathbf{r}_{nlm}\big]},
\end{align}
where
\begin{equation}\label{vnk1}
  V_{nlm,\mathbf{k}\pm}=-\frac{1}{(2\pi)^{3/2}}\int d^3\bm{\rho}\tilde{\phi}^*_{nlm}(\bm{\rho}) \frac{1}{|\bm{\rho}+\mathbf{r}_{nlm}|} e^{\pm i\mathbf{k}\cdot{\bm{\rho}}}.
\end{equation}

Then, we search for those values of the variables $t$, $t'$, and $\mathbf{k}$
that render the action $S_n\pm\mathbf{k}\cdot\mathbf{r}_{nlm}$ stationary,
which yields, in turn,
\begin{align}
  [\mathbf{k}+\mathbf{A}(t')]^2/2 &  = -I_p \label{sadd1}, \\
  [\mathbf{k}+\mathbf{A}(t)]^2/2 &  =A^2(t)/2+E_n \label{sadd2}, \\
  \mathbf{k}_{\textrm{st}}^{\pm}= -\frac{1}{t-t'}\int^t_{t'} & d\tau\mathbf{A}(\tau) \pm \frac{\mathbf{r}_{nlm}}{t-t'} = \mathbf{k}_0 \pm \frac{\mathbf{r}^+_{nlm}}{\tau}, \label{sadd3}
\end{align}
where $\mathbf{k}_0=-\frac{1}{t-t'}\int^t_{t'}d\tau\mathbf{A}(\tau)$
and $ \tau=t-t'$. Equations (\ref{sadd1}) and (\ref{sadd2}) describe,
respectively, energy conservation in the tunneling process and in the
capture process, while Eq.~(\ref{sadd3}) determines the intermediate
electron momentum. The latter takes into account that the electron is
captured at one of the two positions $\pm\mathbf{r}_{nlm}$. The solutions
$(t,t',\mathbf{k})$ of Eqs.~(\ref{sadd1})--(\ref{sadd3}) define the
quantum orbits.

Throughout the paper, we will refer to Eq.~(\ref{amplitude}) as the
quantum model (QM), and the multiple integrals  are performed by numerical
integration with respect to $t$ and $t^{\prime}$ and by saddle-point
integration with respect to $\textbf{k}$. At the end of the pulse, the
population of the $n$th Rydberg state is defined as $P_n=\sum_{l,m}|M_{nlm}|^2$. In our simulation, magnetic
quantum number $m=0$ and (field-free) energy $E_n=-1/(2n^2)$ are adopted.
The ten-cycle laser electric field is $\mathbf{E}(t)=E_{0}\sin\omega
t\mathbf{\hat{e}_z}$ with the vector potential $\mathbf{A}(t)=E_{0}/\omega\cos\omega t\mathbf{\hat{e}_z}$
($\mathbf{\hat{e}_z}$ is a unit polarization vector and the
wavelength $\lambda = 2\pi c/\omega= 800$ nm).  The 1$s$ atomic orbital is expressed as
$\phi_{1s}(\mathbf{r})=\frac{1}{\sqrt{\pi}} \kappa^{3/2}e^{-\kappa
r}$ with $\kappa=\sqrt{2I_{p}}$. For Xe(5$p$), $\phi_{5p}(\mathbf{r})
=\frac{(2\xi)^{11/2}}{\sqrt{10!}}\sqrt{\frac{3}{4\pi}}
r^{4}e^{-\kappa r}\cos\theta$ with $\xi=0.65\sqrt{2I_{p}}$ and
$\kappa=0.94\sqrt{2I_{p}}$. The ionization potential of the H,
He, and Xe atoms are 0.5 a.u., 0.9 a.u., and 0.44 a.u., respectively.
For the TDSE simulation, the details can be found in Ref.
\cite{lin2014PRA}.

\section{Results and discussions}

\textbf{I. Comparison of TDSE and model calculations}\\


\begin{figure}[htb]
\centering\vspace{-0in}\includegraphics[width=3.6 in]{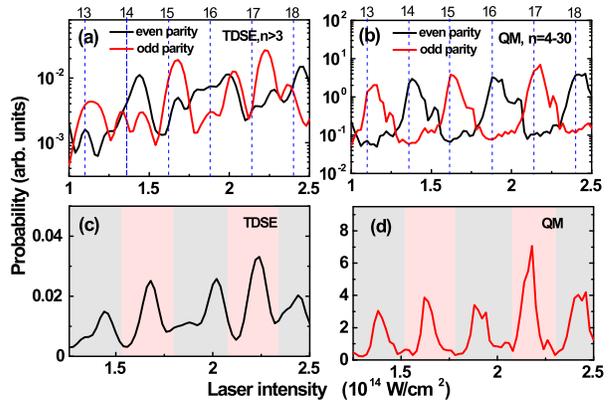}
\caption{ (color online). Populations of Rydberg
states with even or odd parity versus laser intensity calculated
via TDSE (a) and the quantum model (QM)(b) for an initial 1$s$ state. The positions of the $N$-photon channel closings are indicated on the top $x$ axis. For visual convenience, the peaks are separated by dashed lines and underlaid with
alternatingly gray or pink backgrounds. Total Rydberg-state populations on a linear scale (sum over parities)
as functions of the laser intensity simulated by (c): the TDSE and (d): the QM for an initial 1$s$ state.}\label{fig2}%
\end{figure}

In Figs. \ref{fig2}(a) and \ref{fig2}(b), we show the probability
of Rydberg states with principal quantum numbers $n>3$ and with
opposite parities of the final Rydberg states ($l$ even or odd)
calculated by the TDSE and the quantum model for an initial $1s$
state of the H atom as functions of the laser intensity. (For the
TDSE, the pulse envelope is given by a cosine square function with
full width at half maximum of 10 fs, and the details of the
simulation can be found in \cite{lin2014PRA}). Peaks separated by
an intensity interval of about 25\,TW/cm$^{2}$, corresponding to a
shift of the ponderomotive energy $U_p$ by the energy of a laser
photon, i.e., $\Delta U_{p}=\hbar\omega$ ($\approx 1.55$ eV), can
be clearly seen in Fig. \ref{fig2} for both the TDSE and the
quantum model calculation. The results of the quantum model are
almost quantitatively consistent with the TDSE results except for
a small shift of the peaks. In particular, for initial states and
Rydberg states with specified parities, two consecutive peaks are
separated by about $\Delta U_{p}=2\hbar\omega$. These obvious
quantum features can be easily understood in the multiphoton
transition picture of RSE \cite{voklova2010JETP,Freeman1987PRL}.
Since the energies of the Rydberg states are near the threshold so
that their Stark shifts are all close to the Stark shift $U_p$ of
the continuum,  multiphoton resonance between the Rydberg and the
initial states occurs at intensities that are separated by $\Delta
U_{p}=\hbar\omega$ \cite{Freeman1987PRL}. In addition, the dipole
selection rule only allows even-order (odd-order) multiphoton
transitions between states of equal (opposite) parity, which gives
rise to a $\Delta U_{p}=2\hbar\omega$ separation between
consecutive peaks for a transition between two states with
specified parity \cite{lin2014PRA,Piraux2017PRA}. Apparently,
these features are beyond the scope of the semiclassical picture
of the RSE.

Furthermore, closer inspection shows that the peaks alternate in
height as shown in Figs. \ref{fig2}(c) and \ref{fig2}(d), which
display the population of the Rydberg states with both parities.
This can also be observed in the results of Refs.
\cite{voklova2010JETP,lin2014PRA,ZPIE2017} though it was not
addressed there.
This feature is difficult,
if not impossible, to understand in the multiphoton-resonance
picture. For a transition from the ground state to the Rydberg
state, the electron has to absorb more than ten photons under typical
laser conditions as considered here ($\hbar\omega =1.55$ eV and $U_p=6$ eV
for $I=1\times 10^{14}$ W/cm$^2$). The density of the Rydberg states
does not depend on the parity. Hence, no mechanism such as a selection
rule can give rise to a structure that depends on whether an even
or odd number of photons is absorbed in the process.\\

\textbf{II. Analysis of the quantum model}\\

\begin{figure}[htb]
\centering\vspace{-0in}\includegraphics[width=3.6 in]{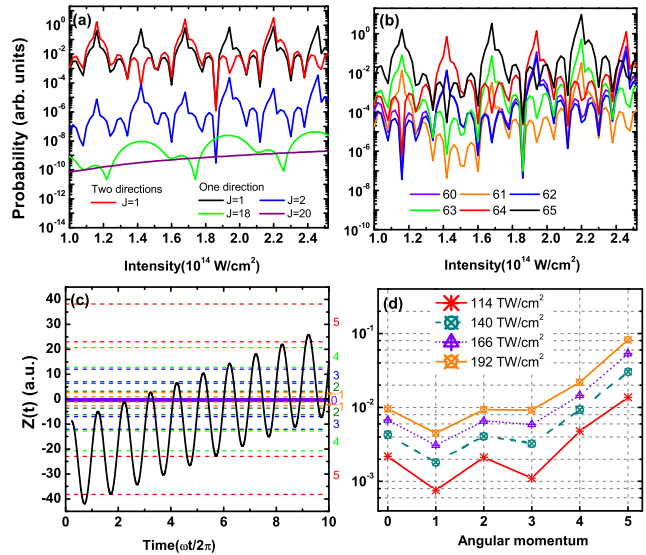}
\caption{ (color online). (a): Interference pattern generated by
orbits with different values of $J$ as given (see text for
details). (b): The population of RSE for different $l$ for a fixed
principal quantum number $n=6$ in an 800-nm laser field. (c): The
electron trajectory in the field $\mathbf{E}(t)=E_0\sin\omega t
\mathbf{e}_z$ corresponding to ionization at the phase $\omega
t_0=92^\circ$\emph{after} the field maximum. The laser intensity
is $1.14\times 10^{14}$ W/cm$^2$ and the transverse velocity is 0
a.u. The colored horizontal dashed lines indicate the boundaries
(in the coordinate $z$) of those regions where the Rydberg states
with $n=6$ and $l\le 5$ have their highest density as shown in
Fig. 8 below. (d): The populations of
specific $l$ within the semiclassical picture for $n=6$ at four
different
peak intensities (see the text for details).}%
\label{fig3}
\end{figure}

In order to elucidate these intriguing features, we classify
the quantum orbits in our model according to the
length of their travel time, that is, the time difference $\Delta t
=t-t'$ between recapture and ionization. Specifically, with the
number $J$ we denote the pair of orbits where $(J-1)T/2<\Delta t
<JT/2$ where $T=2\pi/\omega$. We call an orbit even or odd
according to whether $J$ is even or odd.
Figure \ref{fig3}(a) shows the results calculated using
Eq.~(\ref{amplitude}) for the Rydberg state $n=6$ and $l=5$ for
different electronic orbits. We take the Rydberg state $n=6$,
since for this state the calculated results show a maximum in our
laser intensity range \cite{NISS2009,lin2014PRA}. We notice that
the intensity dependence is a straight line for the orbit with
$J=20$. For $J=18$, an oscillatory structure is beginning to
develop, which becomes more and more apparent when $J$ becomes
smaller (not shown here). Finally, for $J=1$ and 2, pronounced
peaks have emerged. This is because the total pulse duration used
in our calculation is $10\,T$. Namely, for the orbit with $J=20$ the
electron must be freed in the earliest half cycle while for orbits
with smaller $J$ electrons liberated during more and more
subsequent half cycles contribute and interfere with each other,
which results in the peak structure. Hence, the peak structure in
the intensity dependence of the RSE can be attributed to
interference between electron wave packets generated in different
optical half cycles during the laser pulse, which is in agreement
with Ref. \cite{ZPIE2017}.


For a closer look, we consider Fig. \ref{fig1}. For a sufficiently
long (i.e., almost perodic) laser pulse, the ionization dynamics
triggered by this pulse repeat themselves from one cycle to the
next (see, e.g., orbits 1 and 3). Following a similar analysis
in Ref. \cite{Milosev2007}, if a saddle point solving
Eqs.~(\ref{sadd1}) - (\ref{sadd3}) with the Rydberg-state
recombination site in the positive direction [+ sign in
Eq.~(\ref{sadd3})] is given by $(\mathbf{k},t,t')$, then the
saddle point in the next optical cycle with the same Rydberg
state in the same direction is $(\mathbf{k},\,t+T,\, t'+T)$,
and the corresponding action is related by
\begin{equation}
S_n(t+T,t'+T,\mathbf{k})=S_n(t,t',\mathbf{k})+(E_n+I_p+U_p)T.\label{PD1}
\end{equation}
So the contributions of these trajectories differ by the phase
$(U_p+I_p+E_n)T=(U_p+I_p+E_n)2\pi/\omega$ (orbits 1 and 3).
The same holds true for orbits 2 and 4, half a period later when
the field has changed sign.
Due to parity symmetry, a Rydberg state (except for $l=0$)
has two opposite density maxima in the field direction, and the
electron can be captured into one or the other. In the figure,
orbits 1 and 3 go into one direction and orbits 2 and 4 into the
other. If a saddle point solving Eqs.~(\ref{sadd1}) - (\ref{sadd3})
with the Rydberg-state recombination location in the positive direction
[+ sign in Eq.~(\ref{sadd3})] is written as $(\mathbf{k},t,t')$, then the
saddle point solving these equations with the same recombination
location in the negative direction [- sign in Eq.~(\ref{sadd3})] is
$(-\mathbf{k},\,t+T/2,\, t'+T/2)$, and the corresponding action is
given by
\begin{equation}
S_n(t+T/2,t'+T/2,-\mathbf{k})=S_n(t,t',\mathbf{k})+(E_n+I_p+U_p)T/2.\label{PD}
\end{equation}
So the contributions of 2 and 4 differ from those of 1 and 3
by the phase $(U_p+I_p+E_n)T/2=(U_p+I_p+E_n)\pi/\omega$.
The interference of the contributions of orbits 1 and 2 with those of
orbits 3 and 4, i. e., the contributions of two directions, gives
rise to peaks in the RSE separated by $\Delta U_p=2\hbar\omega$,
while the afore-mentioned periodicity from cycle to cycle yields
peaks with the separation of $\Delta U_p=\hbar\omega$, which is
common for all interferences in a field with period $T$. These two
kinds of interference can be clearly seen in Fig. \ref{fig3}(a).

Another element to be considered are the phases due to the
parities of the initial ground state and the final Rydberg state,
which are $(-1)^p$ with $p=\pm 1$ the respective parities. Hence,
if the initial and the final state have opposite parities, an
additional phase of $\pi$ occurs. We notice that the interference
between quantum orbits that are recaptured in opposite directions
of the same Rydberg state plays the same role in the quantum model
as the selection rule does in the multiphoton transition picture.


It is noteworthy that the peaks in Figs. \ref{fig2}(b) and \ref{fig2}(d)
also closely satisfy the channel-closing (CC) condition
$I_p+U_p=m\omega$. For example, the two peaks at $1.16\times
10^{14}$ W/cm$^2$ and $1.42\times 10^{14}$ W/cm$^2$ in Fig.
\ref{fig2}(b) correspond to $U_p=6.9$ and 8.5 eV and, with the
ionization potential $I_p=13.6$ eV and the negligible ionization
potential of the highly excited Rydberg state, satisfy the CC
condition for $m=13$ and $m=14$, in agreement with the Freeman
resonance picture and the selection rule.

From the above analysis, as schematically illustrated in
Fig.~\ref{fig1} and elaborated above, we find that there are
essentially two types of interference of the quantum orbits in the
excitation of a Rydberg state with specific $n$ and $l$ from the
ground state: i) interference of wave packets released during
different optical cycles and captured in one and the same half of
the spatial region of the Rydberg state gives rise to peaks
separated by a laser intensity difference corresponding to $\Delta
U_p=\hbar\omega$; ii) interference of orbits released in adjacent
half cycles (and captured in opposite spatial regions of the
Rydberg state) results in a peak structure with intensity
difference corresponding to $\Delta U_{p}=2\hbar\omega$, and the
peak positions depending on the relative parity of the Rydberg and
the ground states.

Figure \ref{fig3}(b) displays the population of different orbital
angular momenta $l$ for $n=6$.
In the intensity range from $1\times 10^{14}$ W/cm$^2$ to
$2.5\times 10^{14}$ W/cm$^2$ the Rydberg states with $l=4$ and 5
are predominantly populated. The maximal populations of $l=4$ and
5 as a function of laser intensity correspond to the low and high
peaks in Figs. \ref{fig2}(b) and \ref{fig2}(d), respectively.

For a different view of these features, we now turn to a
semiclassical picture (see the Appendix for the
details). In Fig. \ref{fig3}(c), we present an example of an
electron trajectory for  ionization at $\omega t_0=92^\circ$ with
zero initial longitudinal and transverse velocity. We assume that
the probability of capture into the Rydberg state $(n,l,m)$ is
maximal if the electron trajectory $(X(t),Z(t))$ passes the
spatial region where its density $|\phi_{nlm}(\mathbf{r})|^2$ is
highest at a time where its kinetic energy is very low. We
determine this spatial region from a graph of the Rydberg-state
wave function (see Fig. 8 of the Appendix). For the
kinetic energy we require $E_\mathrm{kin}<0.05$ a.u., and there
are two time domains which satisfy the capture condition for every
optical cycle (see Fig. 7 of the Appendix). Figure
\ref{fig3}(d) displays the angular-momentum distributions obtained
this way. Clearly, regardless of intensity $l=5$ dominates
all other angular momenta, 
which is in reasonable agreement with the results of the QM model
in Fig. \ref{fig3}(b). In addition, it can be seen that the
population of $l=5$ for 166 TW/cm$^2$ is higher than that of $l=4$
for 192 TW/cm$^2$ and also for 140 TW/cm$^2$, which confirms the
alternating heights of the peaks in the quantum model displayed by
Fig. \ref{fig3}(b). Therefore, 
the alternating heights of the RSE peaks shown in Fig. \ref{fig2}
are already engrained in this ``simple-man model''. It should be
noted that even though the population at 114 TW/cm$^2$ is higher than
that of 140 TW/cm$^2$ in Figs. \ref{fig3}(b) and \ref{fig3}(d),
this is not the case for the total population shown in Figs.
\ref{fig2}(a) and \ref{fig2}(b). This can be attributed to the fact
that the relative contributions from other Rydberg states with
different $n$ change and affect the modulation when the
intensity is low.\\


\textbf{III. Comparison with experiments}\\

\begin{figure}[htb]
\centering\vspace{-0in}\includegraphics[width=3.6 in]{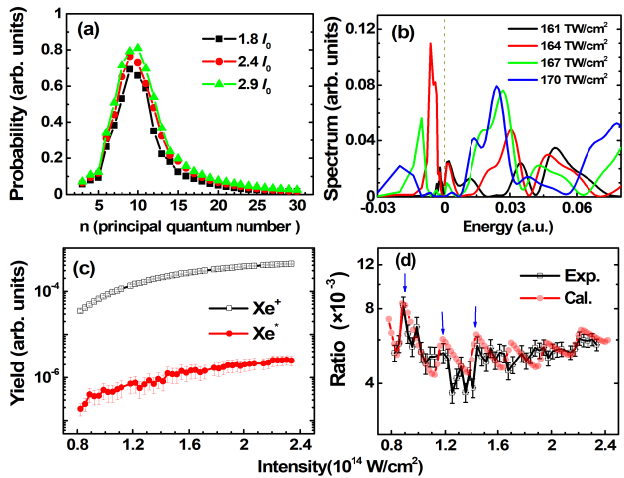}
\caption{ (color online).
(a): \emph{Focal-averaged} RSE vs.  principal quantum number calculated
by the quantum model for the He atom, for $I_0=10^{15}$ W/cm$^2$;
(b): Calculated electron spectra below (multiplied by a factor of 0.2)
and above the continuum threshold for the hydrogen atom and  various fixed intensities;
(c): Experimental yields of single ionization and RSE of the Xe atom;
(d): Experimental and calculated ratios between the Rydberg excitation
and the single ionization yields (see the text for more detail). }%
\label{fig4}
\end{figure}

Experiments show that for the He atom the population of the
Rydberg states peaks at about $n=9$ at $I=1.8\times 10^{15}$
W/cm$^2$ with a slight tendency to shift to higher $n$ when the
intensity increases to $I=2.9\times 10^{15}$ W/cm$^2$
\cite{Eichmann2015PRL}.
For comparison, the focal average of the RSE calculated by our
proposed quantum model is shown in Fig. \ref{fig4}(a) which shows
that the maximum of the RSE shifts to higher energy for increasing intensity, in qualitative
agreement with the experiment. For fixed-intensity RSE
yields, however, our quantum model shows a reverse intensity
dependence (see Fig. \ref{fig4}(b)), which reproduces the results of TDSE
simulations \cite{lin2014PRA}.
This apparent conflict can be resolved as follows: According to
the physical picture underlying our quantum model, the electron
after tunneling out from the ground state oscillates in the laser
field while drifting towards the detector. When it reaches
the position of the Rydberg state, provided its instantaneous
kinetic energy is very low, it can be captured (see the typical
orbits in Fig. \ref{fig1}). Therefore, for small drift momentum,
it can be expected that the RSE yield becomes maximal around
principal quantum numbers $n$ roughly given by $n^2\approx
4\sqrt{U_p}/\omega$, so that the yield slightly shifts with
increasing intensity as shown in Fig. \ref{fig4}(a). The positions
of the maxima of the RSE yields in the fixed-intensity results of
Fig. \ref{fig4}(b) and Ref. \cite{lin2014PRA} are determined by
interference as was shown above. However, interference effects are
largely smeared out by focal averaging, which restores agreement
with the TDSE and the semiclassical calculations in Ref.
\cite{Eichmann2015PRL}. It should be noted that this $n$
distribution cannot be explained by the Freeman resonance
mechanism.

Moreover, we experimentally measured the single-ionization yields
and the RSE of the Xe atom for principal quantum numbers between
$n=20$ and $n=30$. The experimental setup is introduced in Ref.
\cite{Lv2016PRA} (see the Appendix for more
details). The results for Xe are presented in Figs. \ref{fig4}(c)
and \ref{fig4}(d). The intensity-dependent ionization yield
follows a smooth curve. However, the RSE yield does display some
peak structure  especially in the low intensity regime below
$I=1.5\times 10^{14}$ W/cm$^2$ [see Fig. \ref{fig4}(c)]. This is
especially evident in the ratio  Xe$^*$/Xe$^+$, which exhibits
peaks at $I_1=0.89\times 10^{14}$ W/cm$^2$, $I_2=1.18\times
10^{14}$ W/cm$^2$, and $I_3=1.45\times 10^{14}$ W/cm$^2$ [see the
blue arrows in Fig. \ref{fig4}(d)].  This structure is
qualitatively reproduced by the quantum model
if focal averaging is included, indicating that this structure is
the remainder of the peak structure after taking into account the
intensity distribution in the focus.\\

\section{Conclusions and Perspectives}

In summary, we propose a new quantum model based in the spirit of the SFA to
study the RSE process in an intense infrared laser field. In this
quantum model, the electron is first pumped by the laser field
into the continuum where subsequently it evolves in the laser
field. Most of the liberated electrons end up as free electrons
(ATI); however, some may be captured into a Rydberg state. The
electronic wave packets liberated in different optical half cycles
are responsible for peak structures in the intensity dependence of
the RSE yield. The peaks exhibit a modulation, which can be
attributed to a strong dependence of the capture probability on
the spatial position and the parity of the Rydberg state. Our
calculation also well reproduces the experimentally observed
atomic RSE distribution and the intensity dependence of RSE of
atoms. The quantum picture of the RSE in an intense laser field
can be understood as a coherent recapture process accompanying
above-threshold ionization.

We expect similar quantum effects for molecules. For example, RSE
of the O$_2$ molecule is suppressed compared with that of its
companion atom Xe, and this suppression is stronger than the
corresponding suppression of ionization of O$_2$ compared with
that of Xe \cite{Lv2016PRA}. Apparently, this cannot be described
by the semiclassical model. Another such example will be
two-center interference, which has been accepted to be essential
in molecular ionization \cite{Becker2000PRL,lin2012PRL}. With some
extensions, our quantum model  can be applied to investigate these
intriguing phenomena and reveal the underlying physics. Work along
these lines is in progress.\\

\section{Acknowledgment}
The authors acknowledge Xiaojun Liu for helpful discussions. This
work was supported by the National Key program for S\&T Research
and Development (No. 2016YFA0401100), NNSFC (Nos. 11804405,
11425414, and 11534004), and Fundamental Research Fund of Sun Yat-Sen
University (18lgpy77).

\section{APPENDIX:~~Semiclassical analysis and experimental technique}

\textbf{Semiclassical analysis}\\

In the simpleman picture where the ionic Coulomb potential is
ignored, the equation of motion of the electron in the laser field
after ionization is
\begin{equation}\tag{A1}
\ddot{\mathbf{r}}(t)=-\mathbf{E}(t). \label{smm}
\end{equation}
Here the electric field is $\mathbf{E}(t)=E_{0}\sin\omega
t\mathbf{\hat{e}_z}$ with the vector potential
$\mathbf{A}(t)=E_{0}/\omega\cos\omega t\mathbf{\hat{e}_z}$
($\mathbf{\hat{e}_z}$ is a unit polarization vector), and the
wavelength is $\lambda =$ 800 nm. The electron trajectory in the
laser field starts at the tunnel exit $z_0=-I_p/E(t_0)$ with zero
longitudinal and nonzero initial transverse velocity
$\mathbf{v}(t_0)=(0,v_0)$. If we integrate the equation of motion,
we obtain
\begin{align}\tag{A2}
  dz(t)/dt = &  A(t)-A(t_0), \nonumber \\
  dx(t)/dt = &  v_0. \label{velo} \nonumber
\end{align}
Integrating again we get the trajectory of the electron:
\begin{align}\tag{A3}
  z(t) & =z(t_0)-A(t_0)(t-t_0)+\int^t_{t_0}A(\tau)d\tau \nonumber \\
       &=\! -\frac{I_p}{E_0\sin\omega t_0}\!-\!\frac{E_0}{\omega}\cos\omega t_0(t\!-\!t_0)\!+\!\frac{E_0}{\omega^2}(\sin\omega t\!-\!\sin\omega t_0), \nonumber \\
  x(t) &= v_0(t-t_0). \label{traj} \nonumber
\end{align}

The semiclassical physical picture of the RSE can be summarized as
follows: the electron is released at the time $t_0$ into a continuum
Volkov state. Subsequently, it evolves in the laser field and can
be captured into the Rydberg state at time $t$ provided (i) it reaches the
spatial region where the Rydberg state has a high density
 (here we use the condition 
$|\phi|^2 >
0.8|\phi|^2_{\textrm{max}}$) and (ii) its kinetic energy $E_{kin}$ is
small (here we use $E_{kin}\leqslant 0.05$ a.u.). It should
be noted that the result is not sensitive to these criteria.
\begin{figure}[htbp]
  \centering
  \includegraphics[scale=0.35]{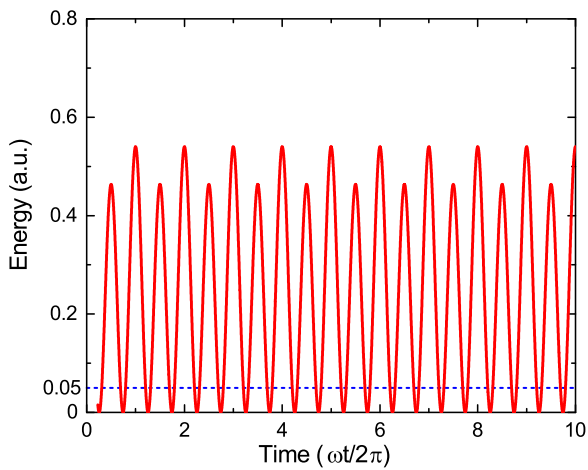}\\
  \caption{Kinetic energy of the electron during its evolution in a laser pulse of 10
  optical cycles. The laser intensity is $1.14\times 10^{14}$ W/cm$^2$, and
  the initial phase and the transverse velocity are 92$^\circ$ and 0, respectively.}\label{fig22}
\end{figure}

In Fig.~\ref{fig22}, we show the evolution of the kinetic energy of
the electron under the  given initial conditions and parameters. It can be seen that there are
two time intervals in each optical cycle where the kinetic
energy is small enough for capture, i. e., $E_{kin}\leqslant 0.05$ a.u.,
as defined above.

\begin{figure}[htbp]
  \centering
  \includegraphics[scale=0.3]{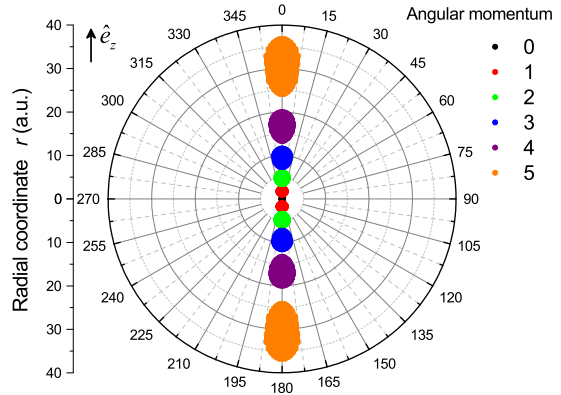}\\
  \caption{The regions in polar coordinates where the Rydberg-state densities
$|\phi_{nlm}(\mathbf{r})|^2$ are larger than 80 percent of their
maximal values for $n=6$  and angular momenta $l\le n-1$.
Different colors denote
different angular momenta.}\label{fig44}
\end{figure}

For Rydberg states of $n=6$ and different values of $l$
(see Fig. \ref{fig33}),
Fig. \ref{fig44} identifies for these $l$ the regions  where the
density criterion $|\phi|^2 > 0.8|\phi|^2_{\textrm{max}}$ is
satisfied. Clearly, the relevant region moves towards larger
distances from the origin with increasing $l$.

In our calculation, each electron trajectory is released via
tunneling at some instant during the laser-pulse duration (a
10-cycle pulse with constant electric amplitude is considered).
Its weight, which  is dependent on the ionization rate and the
initial transverse velocity (from -0.3 a.u. to 0.3 a.u.), is
determined as in Ref.~\cite{NISS2009}. Then the electron evolves
in the laser field until the end of the laser pulse. If it reaches
the regions shown in Fig.~\ref{fig44} and its kinetic energy is
smaller than 0.05 a.u., it is considered to be captured by the
Rydberg state $\phi_{nlm}$. The final Rydberg excitation
probabilities shown in Fig. \ref{fig3}(d) of the main body of the paper are
obtained from statistic of all the trajectories satisfying the
afore-mentioned criteria of capture (about 10$^4$-10$^6$
trajectories for each state depending on $l$) in overall
2$\times10^7$ trajectories.\\


\textbf{Experimental technique}\\

In our experiments, we applied the delayed-static-field-ionization
method to ionize the neutral Rydbergs, using a time-of-flight
(ToF) mass spectrometer operated under a pulsed-electric-field
mode. Experimentally, an effusive atomic or molecular beam through
a leak valve interacted with a focused Ti:Sapphire femtosecond
laser with a central wavelength of 800 nm and pulse duration of 50
fs. After the direct-ionized ions (Xe$^+$) were pushed away from
the detector by an electric field, the remaining high-lying
neutral Rydbergs (Xe$^*$) were field-ionized by another electric
field with a delay time of typically 1.0 $\mu$s, and were detected
by dual micro-channel plates at the end of the flight section of
about 50 cm. In the case of detection of Xe$^+$, standard dc
electric fields were applied in the ToF mass spectrometer. The
voltages in both cases were kept the same to ensure identical
detection efficiencies for Xe$^+$ and (Xe$^*$)$^+$. This allows us
to detect the neutral Rydberg states with $20<n<30$, estimated by
the  saddle-point model of static-field ionization $[F=1/(9n^4)]$.
The laser-pulse energy was controlled by a half-wave plate and a
Glan prism, before being focused into the vacuum chamber with a
250 mm lens. The peak intensity of the focused laser pulse was
calibrated by comparing the measured saturation intensity of Xe
with that calculated by the ADK model \cite{ADK}. The scanning of the
laser intensity was precisely controlled by simultaneously
monitoring the pulse energy using a fast photodiode while rotating
the half-wave plate. Each data point was an averaged result of
$10^4$ laser shots with an intensity uncertainty of 1 TW/cm$^2$.\\


\end{document}